\documentclass[journal]{IEEEtran}

\usepackage{graphicx}
\usepackage{cite}
\usepackage{xfrac}
\usepackage{amsmath}
\usepackage{mathtools}
\usepackage{xcolor}

\begin{document}

\title{Comparing Different Preprocessing Methods in Automated Segmentation of Retinal Vasculature}
%
%
\author{Meysam Tavakoli,
        Faraz Kalantari, 
        Alireza Golestaneh,
\thanks{M. Tavakoli is with the Dept. of Physics, Indiana University-Purdue University, Indianapolis, IN, USA.}
\thanks{A. Golestaneh is with the Electrical Engineering Department, Arizona State University, Tempe, AZ, USA}
\thanks{F. Kalantari is with Department of Radiation Oncology, University of Texas Southwestern Medical Center, Dallas, TX, USA
}
\\
\textbf{To appear in: 2017 IEEE Nuclear Science Symposium and Medical Imaging Conference (NSS/MIC) \\
DOI: 10.1109/NSSMIC.2017.8532607}
}
\maketitle

\begin{abstract}
Computer methods and image processing provide medical doctors  assistance at any time and relieve their work load, especially for iterative processes like identifying objects of interest such as lesions and anatomical structures from the image. Vescular detection is considered to be a crucial step in some retinal image analysis algorithms to find other retinal landmarks and lesions, and their corresponding diameters, to use as a length reference to measure objects in the retina.
The objective of this study is to compare effect of two preprocessing methods on retinal vessel segmentation methods, Laplacian-of-Gaussian edge detector (using second-order spatial differentiation), Canny edge detector (estimating the gradient intensity), and Matched filter edge detector either in the normal fundus images or in the presence of retinal lesions like diabetic retinopathy. The steps for the segmentation are as following: 1) Smoothing: suppress as much noise as possible, without destroying the true edges, 2) Enhancement: apply a filter to enhance the quality of the edges in the image (sharpening), 3) Detection: determine which edge pixels should be discarded as noise and which should be retained by thresholding the edge strength and edge size, 4) Localization: determine the exact location of an edge by edge thinning or linking.
From the accuracy view point, comparing to manual segmentation performed by ophthalmologists for retinal images belonging to a test set of 120 images, by using first preprocessing method, Illumination equalization, and contrast enhancement
, the accuracy of Canny, Laplacian-of-Gaussian, and Match filter vessel segmentation was more than 85\% for all databases (MUMS-DB, DRIVE, MESSIDOR). The performance of the segmentation methods using top-hat preprocessing (the second method) was more than 80\%. And lastly, using matched filter had maximum accuracy for the vessel segmentation for all preprocessing steps for all databases.

\end{abstract}


%
\IEEEpeerreviewmaketitle

\section{Introduction}
\IEEEPARstart{T}{he} transparent living tissue of eye makes retina to be the only part of body where vascular network is directly visible. Many systemic diseases change the vascular network and could be diagnosed through this transparent window.
Evaluation of retinal vascular network to find abnormality was done by ophthalmologists, which is time consuming and associated with error and fatigue. Moreover, clinical analysis is based on clinicians' idea may not be repeatable. One possible solution for these problems is to use Computer Assisted Diagnosis (CAD) systems \cite{Khansari-TMI, TavakoliFA, Tavakoli-SPECT}. 
Image processing and computer vision techniques are required to extract suitable information about vascular tree and its alteration. Vessel segmentation is an essential step in some practical applications such as detection of vessels pathology like stenosis or aneurysm, medical image registration. It is also a corner-stone in detection of other retinal land marks such as fovea and Optic Nerve Head (ONH) \cite{Niemeijer1, sinthanayothin}. Moreover, it is critical step in screening of diabetic retinopathy (DR)\cite{Mehdizadeh-color, Gagnon L, Sherwani SM, Muangnak N, Constante P, Khansari-DR2}.\\
Before vessel segmentation in fundus image, the image has to be preprocessed to ensure adequate level of success in detection. Here we applied two different methods (Illumination equalization, and contrast enhancement and top-hat transform) separately and compare the results of vessel segmentation in each of these two ways.\\
The retinal blood vessel network has a number of characteristics that can be used in
segmentation methods: (1) Gaussian shape of the vessel cross-sectional grey level profile. (2) The vessels are piecewise linear, (3) They are continuous in other words; the vessel direction and grey level do not
change suddenly, (4) All vessels originate from a single area, ONH, and they are
connected to all other vessels \cite{Heneghan, Khansari-optics, Fraz1}.
But some factors that obstruct vessel segmentation are:\\
- Size, shape or color of vessels are obviously varied \\
- Some techniques are confused in points that are vessel crossings and
bifurcations \\
- In some methods the edge of the ONH can be incorrectly selected as a
vessel \cite{Tavakoli-ONH1, tavakoli-twopreprocessingsteps, Morales S, Abdullah M}\\
Generally, vessel detection algorithms can be separated to two classes: those based on
pixel base operator such as morphological operators and those based on tracking vessels that start from a point on the vessel after that process the local region and track
the vessel through the image.


 




\section{Related Works}
There are several algorithms for retinal vessel segmentation \cite{sinthanayothin, Heneghan, Sinthanayothin2, Hoover1, Hoover2, Staal, Soares, Jiang X, Wu D, Tolias, Zana, Gardner, Walter, Tavakoli-ONH2}. The studies about pixel based operators are very different, some of that described by
using convolution methods \cite{Sinthanayothin2}, threshold probing method \cite{Heneghan}, morphological
operators \cite{Hoover1, Hoover2, Staal, Soares, Jiang X, Wu D}, wavelets \cite{Tolias}, region growing \cite{Martinez-Perez2} and artificial intelligence used in
pixel classification \cite{sinthanayothin, Zana, Gardner}.
In the tracking or tracing methods the vessel width and orientation were computed in a
local region near the current point after that, the procedure is repeated in the direction
of vessel till vessel is tracked \cite{Can, Vlachos}. \\
On the other had, retinal vessel segmentation methods divided in two methods supervised and
unsupervised techniques \cite{Staal}.
Surprised techniques that are most recent are included: neural networks \cite{Sinthanayothin2}, the K-
nearest neighbour classifier \cite{Staal}, or the Bayesian classifier \cite{Soares}. And Unsupervised
methods that include: matched filtering \cite{Hoover1}, adaptive thresholding \cite{Jiang X, Wu D}, vessel
tracking \cite{Wu D, Tolias, Zana}, and morphological-operators techniques \cite{Zana}.
In the other hand, some studies applied the combination of these two methods for
instance Gardner \cite{Gardner}, Sinthanayothin \cite{sinthanayothin}, or Goldbaum \cite{Goldbaum} used combination of
match filter with artificial neural networks for detecting retinal vessel.

\section{Method}

\begin{figure}[h!]
\includegraphics[width=\linewidth]{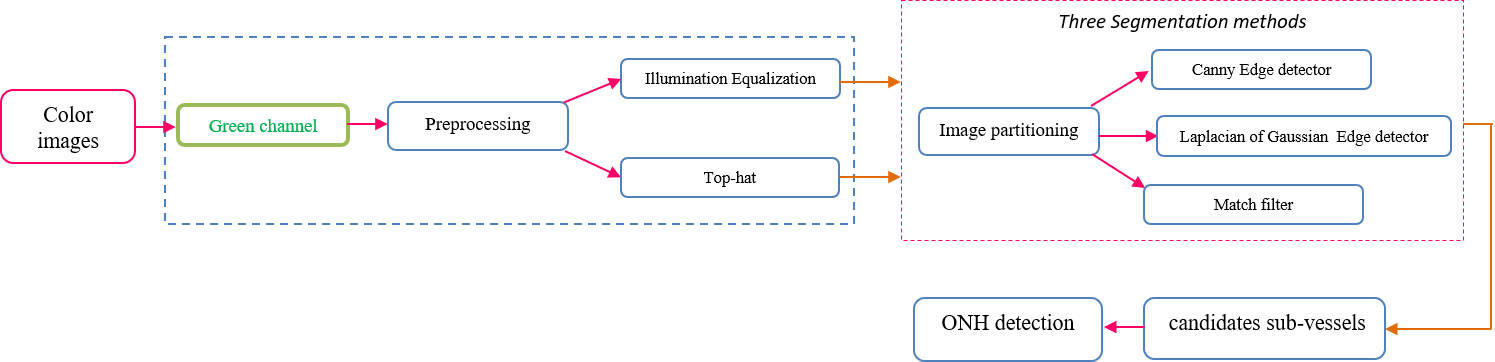}
\centering
\caption{The scheme of the methods used in this study.}
\label{scheme}
\end{figure}

The overall scheme of the methods used in this study is shown in Fig.\ref{scheme}.

\subsection{Materials}

For the vessel segmentation, three databsets, one rural and two publicly available databases, were used. The first rural database was named  Mashhad University Medical Science Database (MUMS-DB). A total of 120 color images were captured which 100 were with DR and 20 normal from both left and right eyes of patientsl. The train set consists of 20 images and our test set consists of 100 images (80 images with DR and 20 images without DR). Images were taken with
a field of view (FOV) of 50° under the same lighting conditions. According to the National
Screening Committee standards all the images are captured by using a TOPCON
(TRC-50EX), 50° Mydriatic retinal camera.
The acquired image resolution is $2896\times1944$ TIFF format  \cite{tavakoli-fluoresceinangiography, tavakoli-radon}. The second database, DRIVE, consisting of 40 images with image resolution of $768\times584$ pixels in which 33 cases did not have any sign of DR and 7 ones showed signs of early or mild DR with a 45° FOV. This database is divided into two sets: testing and training set, each of them containing 20 images \cite{researchsection}. The last database, MESSIDOR dataset includes 120 retinal images of the posterior pole acquired by 3 ophthalmologic departments using a color video 3CCD camera on a Topcon TRC NW6 non-mydriatic retinograph with a 45 degree FOV. The images resolution were captured using 8 bits per color plane at $1440\times960$, $2240\times1488$ or $2304\times1536$ pixels \cite{MESSIDOR}

\subsection{Preprocessing}

 The preprocessing step provides us an image with high possible vessel and background contrast and also unifies the histogram of the images. Here we applied two different methods separately \cite{tavakoli-twopreprocessingsteps} and compare the results of vessel segmentation in each of these two ways: \\
 
\subsubsection{Illumination Equalization, and Contrast Enhancement} 
In Illumination equalization, and contrast enhancement, our purpose is to unify the histogram of all fundus images. In background elimination, background of the image is filtered to some level and we process just the foreground image. For this preprocessing level, there are three steps that described in this study: color space conversion, illumination equalization, and contrast enhancement \cite{tavakoli-twopreprocessingsteps}.
At the first step the RGB fundus image is converted to the HSV color system. In the second step, we search for retinal region detection, the region of the retina from the HSV image. In the next step, the  Illumination equalization, the original RGB image takes as an input and equalizes uneven illumination in the image. The output of the part represents the channels of the original image where the illumination has been equalized. 
At last, in contrast enhancement we determine an image selected as a reference image and use its color histogram as template for all fundus images to normalize the background then in each of sub-images contrast enhancement were done. In better words, since different retinal images have different brightness, this approach  helps us to use the same threshold for all fundus images. For this purpose, a reference image is selected and the threshold is adjusted. Then histogram specification is used to incorporate the images.
With performing preprocessing, we have an image with maximum possible contrast between the retinal features and background and also unify the histogram of the available fundus images.

\subsubsection{Top-hat transformation}
 Although retinal images have three components (R, G, B), their green channel has the best contrast between vessel and background; so the green channel is selected as input image.  
Here, we also used mathematical morphology operators.  Morphological operations work with two parts. The first one is the image to be processed and the second is called structure element. 
Erosion is used to reduce the objects in the image with the structure element, also known as the kernel. In contrast, dilation is used to increase the objects in the image. Secondary operations that depend on erosion and dilation are opening and closing operations. Opening, denoted as $f \circ b$, is applying an erosion followed by a dilation operation. The b represents the structure element. On the other hand, closing is first applying dilation then erosion. It is denoted as $f \bullet b$. Building from opening and closing operations, the top-hat transform is defined as the difference between the input image and its opening or closing.
The top-hat transform is one of the important morphological operators \cite{tavakoli-radon}. Based on dilation and erosion, opening and closing denoted by $f \circ b$ and $f \bullet b$ are defined. The top-hat transform is defined as the difference between the input image and its opening. The top-hat transform includes white top-hat transform (WTH) and black top-hat transform (BTH) are defined by:

\begin{equation} \label{eq:sensitivity}
\begin{cases}
WTH(x,y)= f(x,y) - f \circ b(x,y) \\
BTH(x,y)= f \bullet b(x,y) - f(x,y) \\
\end{cases}
\end{equation}

\begin{figure}[h!]
\includegraphics[width=\linewidth]{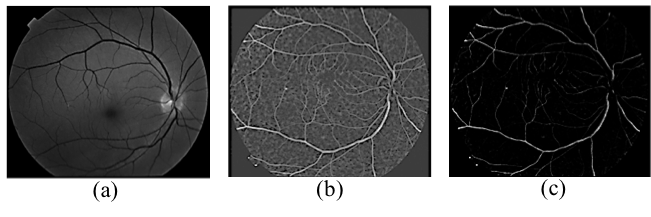}
\centering
\caption{(a) Fundus image from MUMS-DB (b) top-hat result and contrast stretching (c) result of subtraction of top-hat and filtered top-hat image.}
\label{fig2}
\end{figure}

In our preprocessing the basic idea is increasing the contrast between the vessels and background regions of the image. WTH or BTH extract bright and dim image regions corresponding to the used structure element. So, using the concept of WTH or BTH is one way of image enhancement through contrast enlarging based on top-hat transform.
In the fundus images, the background brightness is not the same in the whole image. This background variation would lead to missed vessels or false vessel detection in the following steps. Moreover, in I, background is brighter than the details, however the vessels and other components are preferred to appear brighter than background. To deal with the problem, I is inverted as shown in I= 255 - I.
Since we need a uniform background, to decrease the intensity variations in vessels background, we were firstly applied WTH(x,y) on image. It gave a high degree of differentiation between these features and background. A top-hat transformation was based on a ‘‘disk structure element’’ whose diameter was empirically found that the best compromise between the features and background. The disk diameter depended on the input image resolution.
After top-hat transformation, we used contrast stretching to change the contrast or brightness of an image. The result was a linear mapping of a subset of pixel values to the entire range of grays, from the black to the white, producing an image with much higher contrast. Filtering a region is the process of applying a filter to a region of interest in an image,
where a binary mask defines the region.
In this situation we used an averaging filter on the result of image from last section
(top-hat result) after that, we subtract the image from last section with the result of
applying averaging filter. Before this section, in image result from top-hat transform,
there are some variations in the image and some points like noise that without
eliminate these points maybe supposed as some part of vessels, that increase false
positive rate of our algorithm and after applying filter and subtraction this variation was removed. In other words, this section is applied for better removing background
variation for better detecting of vessels \cite{tavakoli-fluoresceinangiography}. The result of first step is shown in Fig. \ref{fig2}. 

\subsection{Procedure of Vessel Detection}
Blood vessels can be described as bright curvilinear objects opposite of a darker
background with unclear edges.
The retinal blood vessels are non-uniform in intensity, length and width throughout
the image. Global techniques that attempt detection over the entire image are not
effective when dealing with blood vessels. Our approach  addresses the image locally and regionally where homogeneity of the vessel is more likely to happen. The algorithm is composed of 3 steps: Generation of sub-images, vessel segmentation, which three segmentation approaches, 1) Laplacian of Gaussian, 2) Canny, and 3) Matched filter used in this study. 
In order to extract ONH, it should be extracted in local windows.
\subsubsection{Multi-overlapping window}
In the proposed algorithm, each fundus image was partitioned into overlapping widows in the first step. To find objects on
border of sub-images, an overlapping pattern of sliding windows was defined. For determining the size of each sub-image or
sliding window our knowledge database was used. In this regard, minimum and maximum sizes of targeted object specify the
size of the windows (n).
The $\textit{n}$ has a direct effect on the extraction accuracy. Another important parameter which affects the algorithm's accuracy is the windows overlapping. 
If window's step is equal to one, we will search every pixel of images just one time and sub-images only touch each other
without any overlapping and if step is defined as two or more then we go (n/step) pixel each time either in horizontal and
vertical sliding then each pixel would belong to up to $n^2$ sub-images \cite{tavakoli-fluoresceinangiography}. In Fig. \ref{fig3} we have shown some sample sub-images in a retinal image.

\begin{figure}[h!]
\includegraphics[width=2in]{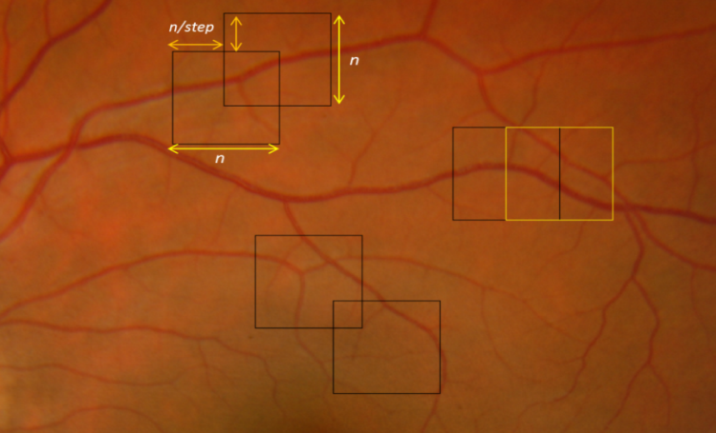}
\centering
\caption{Window size and overlapping ratio.}
\label{fig3}
\end{figure}

\subsubsection{Vessel Segmentation}
Here we applied three retinal vessel segmentation methods, 1) Laplacian of Gaussian (LoG) edge detector \cite{Lowell J}, 2) Canny edge detector \cite{Canny, Zana}, and 3) Matched filter edge detector \cite{Chaudhuri} for detection of ONH either in normal fundus images or in presence of retinal lesion like in diabetic retinopathy (DR). 
In general, the steps for the edge detection are in following: (1) Smoothing: suppress as much noise as possible, without destroying the true edges, (2) Enhancement: apply a filter to enhance the quality of the edges in the image (sharpening), (3) Detection: determine which edge pixels should be discarded as noise and which should be retained by thresholding the edge strength and edge size, (4)	Localization: determine the exact location of an edge by edge thinning or linking.
In the LoG edge detector uses the second-order spatial differentiation.
\begin{equation}
\bigtriangledown ^2 f = \dfrac{\partial^2 f}{\partial x^2} + \dfrac{\partial^2 f}{\partial y^2}
\end{equation}
The Laplacian is usually combined with smoothing as a precursor to finding edges via zero-crossings. The 2-D Gaussian function:
\begin{equation}
h(x,y) = e^{\dfrac{-(x^2 + y^2)}{2 \sigma ^2}}
\end{equation}
Where $\sigma$ is the standard deviation, blurs the image with the degree of blurring being determined by the value of $\sigma$. If an image is pre-smoothed by a Gaussian filter, then we have the LoG operation that is defined:
\begin{equation}
(\bigtriangledown ^2 G_{\sigma})**\textit{I}
\end{equation}
Where $\bigtriangledown ^2 G_{\sigma}(x,y)= (\dfrac{1}{2\pi \sigma^4})(\dfrac{x^2 + y^2}{2 \sigma ^2} - 2)e^{\dfrac{-(x^2 + y^2)}{2 \sigma ^2}}$ \\
In Canny edge detection, we estimate the gradient magnitude, and use this estimate to determine the edge positions and directions. 
\begin{equation}
\begin{cases}
f_{x} = \dfrac{\partial f}{\partial x} = K_{\bigtriangledown_{x}} ** (G_{\sigma}**\textit{I}) = (\bigtriangledown_{x}G_{\sigma})**\textit{I}\\
f_{y} = \dfrac{\partial f}{\partial y} = K_{\bigtriangledown_{y}} ** (G_{\sigma}**\textit{I}) = (\bigtriangledown_{y}G_{\sigma})**\textit{I}
\end{cases}
\end{equation}
Where 
\begin{equation}
\begin{cases}
\bigtriangledown_{x}G_{\sigma}=(\dfrac{-x}{2\pi \sigma^4})e^{\dfrac{-(x^2 + y^2)}{2 \sigma ^2}} \\\bigtriangledown_{y}G_{\sigma}=(\dfrac{-y}{2\pi \sigma^4})e^{\dfrac{-(x^2 + y^2)}{2 \sigma ^2}}
\end{cases}
\end{equation}

The algorithm runs in 4 separate steps: (1) Smooth image with a Gaussian: optimizes the trade-off between noise filtering and edge localization, (2) Compute the Gradient magnitude using approximations of partial derivatives, (3) Thin edges by applying non-maxima suppression to the gradient magnitude, and (4) Detect edges by double thresholding. We can compute the magnitude and orientation of the gradient for each pixel based two filtered images. \\
$|\bigtriangledown f(x,y)|= \surd f_{x}^2 + f_{y}^2$ = rate of change of f(x,y)\\
$\angle \bigtriangledown f(x,y) = tan^-1(\dfrac{f_{y}}{f_{x}})$= orientqtion of rate of f(x,y)

The matched filter has been widely used in the detection of blood vessels of the human retina digital image. In this paper, the matched filter response to the detection of blood vessels is increased by proposing better filter parameters.
The Matched Filter was first proposed in to detect vessels in retinal images. It makes use of the prior knowledge that the cross-section of the vessels can be approximated by a Gaussian function. Therefore, a Gaussian-shaped filter can be used to “match” the vessels for detection. The Matched Filter is defined as
\begin{equation}
G(x,y)=(\dfrac{1}{\surd 2\pi \sigma^2})e^{\dfrac{-(x^2 + y^2)}{2 \sigma ^2}} - m_{0}
\end{equation}
Where $m_{0}$ is chosen to make kernel G(x,y) has zero mean.\\
.

\begin{figure}[h!]
\includegraphics[width=\linewidth]{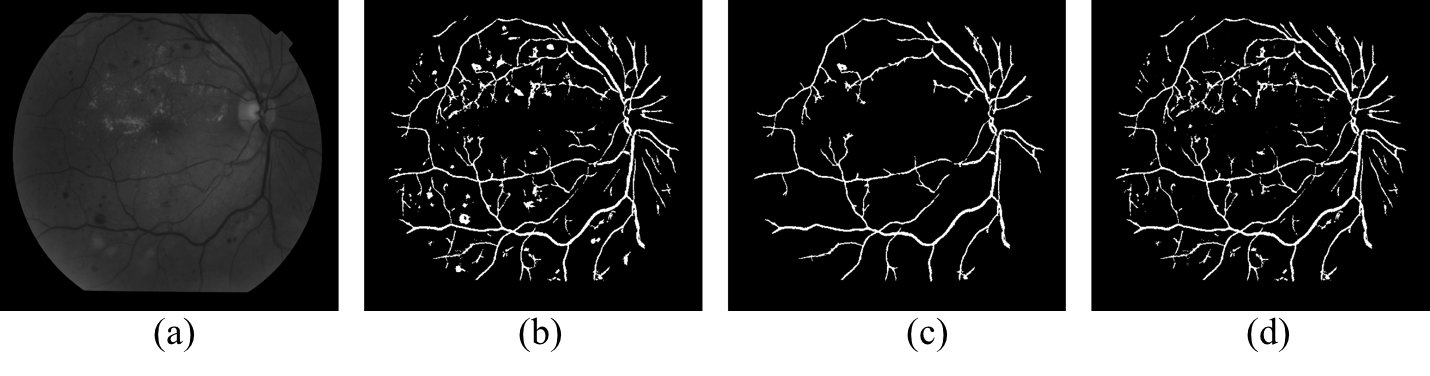}
\centering
\caption{Using illumination equalization, and contrast enhancement preprocessing. Original Green channel fundus image from local database (MUMS-DB), (a) result of segmentation with LoG (b) result of segmentation with Canny (c) result of segmentation with Match filter (d).}
\label{fig3}
\end{figure}

\section{Experimental Results}

To calculate the efficiency of the current methods in retinal vessel segmentation and also to compare the results with other reported studies, it is necessary to compare all pixels of the final automated segmentated images with the manual segmentation or grand truth (GT) files. For the evaluation, we used the concept of sensitivity and specificity, and receiver operating characteristic (ROC) curves.
At first, the algorithm was evaluated in terms of true positive rate (TPR) given by the sensitivity, and false positive rate (FPR), given by specificity. Also the accuracy was determined as a measurement providing the ratio of well-classiﬁed pixels. The results for the automated method compared to the GS were calculated for each image. The higher the sensitivity and specificity values, the better the procedure. Regarding these calculations the proper metrics are defined as \cite{tavakoli-fluoresceinangiography}: 

\begin{equation} \label{eq:sensitivity}
\begin{cases}
Sensitivity = \frac{TP}{TP+FN} \\
Specificity = \frac{TN}{TN+FP} \\
Accuracy = Acc = \frac{TP+TN}{TP+FN+TN+FP}
\end{cases}
\end{equation}

Where TP is true positive, TN is true negative, FP is false positive and FN is false negative.

\subsection{Training and Test Set for the Image Database}
\begin{figure}[h!]
\includegraphics[width=\linewidth]{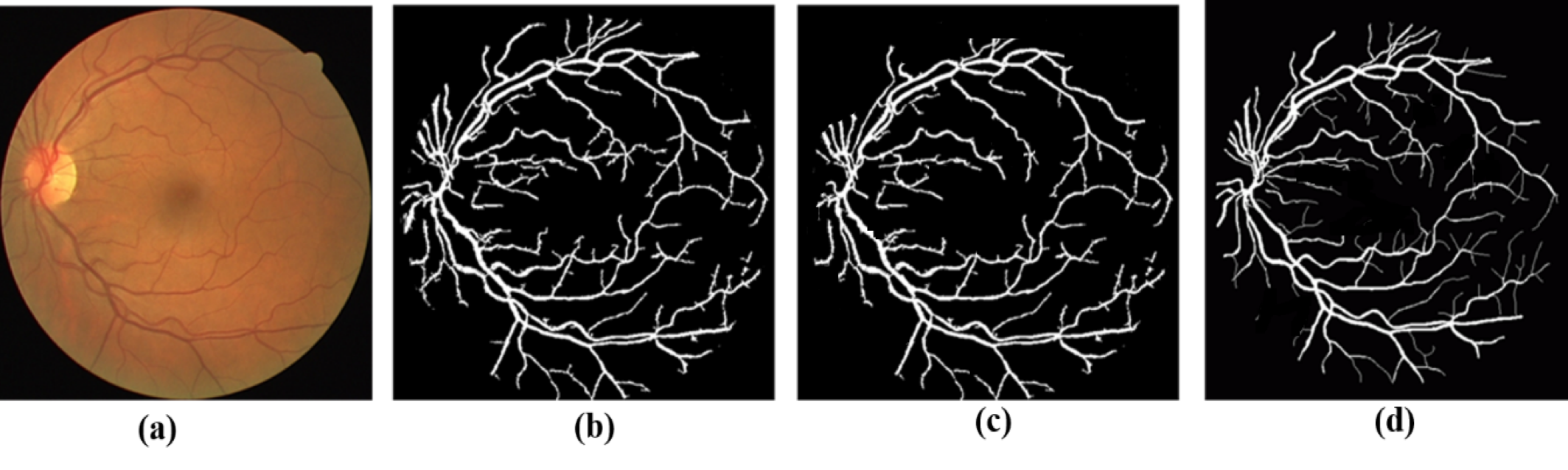}
\centering
\caption{Using illumination equalization, and contrast enhancement preprocessing: Original image (DRIVE) (a) result of segmentation with LoG (b) result of segmentation with Canny (c) result of segmentation with Match filter (d).}
\label{fig4}
\end{figure}

\begin{figure}[h!]
\includegraphics[width=\linewidth]{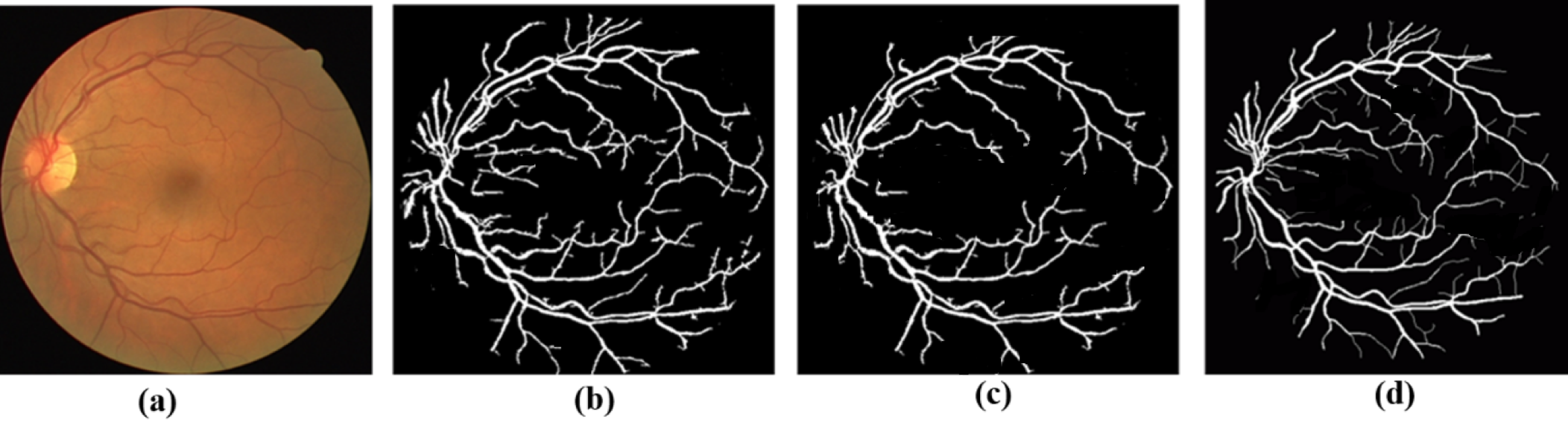}
\centering
\caption{Using top-hat preprocessing: Original image (DRIVE) (a) result of segmentation with LoG (b) result of segmentation with Canny (c) result of segmentation with Match filter (d).}
\label{fig5}
\end{figure}

In this study, we used 60 images for a training set (learning purpose). This consisted of 20 images from all databases (MUMS-DB,  DRIVE, and MESSIDOR). The test set (for the test purposes) consisted of 220 fundus images of which 100 for MUMS-DB, 20 for the DRIVE, and 100 images from MESSIDOR. After setting up the parameters of our algorithm using training set, the methods were tested in each image of the databases in test set.  Some results are shown in Fig. \ref{fig3},  Fig. \ref{fig4}, and Fig. \ref{fig5} from MUMS-DB and DRIVE which compare the results of the vascular segmentation using the two preprocessing.

\subsection{Comparing the Statistics Results of Vessel Segmentation in Three Databases}
The sensitivity of the threshold was also characterized along the Equation (8). A ROC curve is a plot of TPF (Se) versus FPF (1-Sp). 
A ROC curve, plotted to show the effect of a varying threshold, shows the presence or absence of sub-vessels in each sub-image, denoted by the $Th$ parameter (predefined threshold), in our datasets. To plot ROC curve for all of three databases, the value $Th$ is varied over the range [0, 5].  Parameters used for plotting ROC are shown in Table \ref{table2}.

\begin{figure}[h!]
\includegraphics[width=\linewidth]{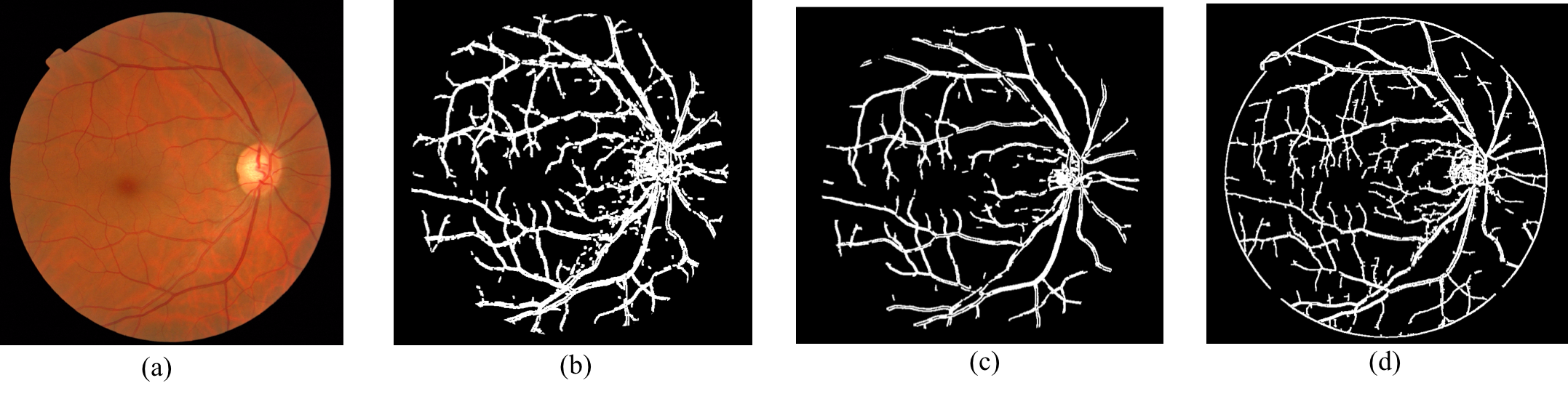}
\centering
\caption{Using top-hat preprocessing: Original image (DRIVE) (a) result of segmentation with LoG (b) result of segmentation with Canny (c) result of segmentation with Match filter (d).}
\label{fig6}
\end{figure}

\begin{figure}[h!]
\includegraphics[width=\linewidth]{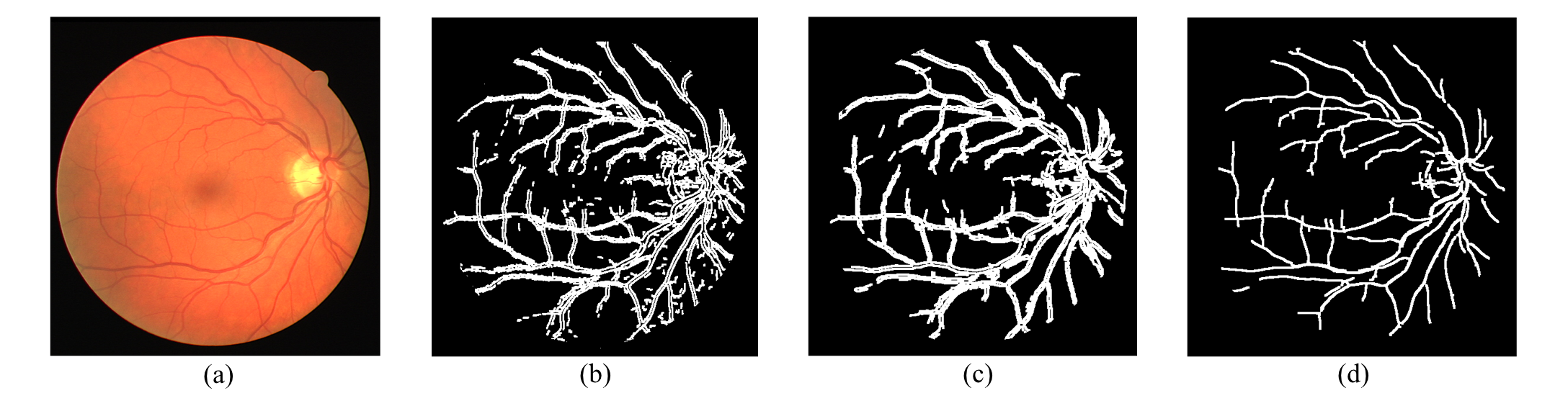}
\centering
\caption{Using illumination equalization, and contrast enhancement preprocessing: Original image (DRIVE) (a) result of segmentation with LoG (b) result of segmentation with Canny (c) result of segmentation with Match filter (d).}
\label{fig7}
\end{figure}

\begin{table}[h!]
\renewcommand{\arraystretch}{1.3}
\caption{PARAMETERS USED IN OUR APPROACH FOR ALL THE THREE  DATABASES}
\label{table2}
\centering
\resizebox{\columnwidth}{!} {
\begin{tabular}{|c|c|c|c|c|c|}
\hline
Database & No. of Images & Window Size (n) & Step & Th \\
\hline\hline
MUMS-DB & 100 & 62 & 5 & [0,5] \\
\hline
DRIVE & 20 & 15 & 6 & [0,5] \\
\hline 
MESSIDOR & 100 & 40 & 5 & [0,5] \\
\hline
\end{tabular}
}
\end{table}
Some results have shown in Figs.~\ref{fig6} and \ref{fig7} for normal image and Fig.~\ref{fig3} for DR image.

All images process with our algorithm and the TPR is achieved by ratio of number of
true pixels hand labeled as blood vessel to total number of hand labeled pixels. Our
result show more than 90\% for TPR.

Statistical information about the sensitivity and specificity measures is extracted. The higher the sensitivity and specificity values, the better the procedure. For all retinal images of test set (220 images), our reader labeled the vessels on the images and the result of this manual detections were saved to be analyzed further. 

By using the first preprocessing, Illumination along with contrast enhancement, for Laplacian-of-Gaussian, Canny, and Match filter our segmentation results were 89.12\%, 85.97\%, and 90.54\% respectively. Also by using the second preprocessing, the top-hat, for LoG, Canny, and Match filter our segmentation results were 83.09\%, 81.17\%, and 89.04\% respectively.

According to manual segmentation, and without considering the preprocessing step, using the Laplacian-of-Gaussian vessel segmentation our automated algorithm finds more than 80\%, 88.5\%, and 87.25\% accuracy for MUMS-DB, DRIVE, and MESSIDOR database.
The Canny vessel segmentation our automated algorithm found 75.40\% in MESSIDOR database and 79.23\% in DRIVE database, and 70.12\% in the local database, MUMS-DB.
At last, using Matched filter in the vessel segmentation our algorithm found with accuracy of 82\%  in MESSIDOR in DRIVE database 87.4\%. For the local database, MUMS-DB, the method segmented  correctly in 82.75\% of all fundus images. \\

\section{Discussion and Conclusion}

Since fundus images are nowadays in the digital format, it is possible to create a
computer-based system that automatically detects landmarks from fundus images \cite{welikala, welikala2, Khansari-DR}. An
automatic system would save the working time of well-paid clinicians, and letting
hospitals to use their valuable resources in other important tasks. It could also be possible to check more people and more often with the help of an automatic
screening system, since it would be more inexpensive than screening by humans.
In this study, we detect retinal blood vessel network without interference of any
ophthalmologist by using automated algorithm system. The performance on the
algorithm was acceptable for vessel detection.
Computers are appropriate to problems involving the derivation of quantitative
information from images because of their capacity to process data in fast and efficient
manner with a high degree of reproducibility \cite{Kochner, Matsopoulos}.
The detection and removal of the retinal blood vessel can facilitate detection of
features that are sign of DR such as Microaneurysms (MAs), Hemorrages (HEs), and any changes in blood vessels like
neovascularization and venous changes.
As well as another application for retinal vascular detection technique is for
recognizing the identity of persons \cite{Martinez-Perez} because vascular network are as the
fingerprint in identification.
Small vessels can appear as small isolated patterns, which can be missed in detection
procedure.
Vascular detection in the first step of main processing, removes the vascular tree map
from the retina.
For usual image, in our algorithm, it is common to accept that nearly 90\% of vessel
pixels can be detected. Gardner et al \cite{Gardner} by using ANN reached to a sensitivity and
specificity of 91\% for retinal vessel segmentation in red free images that is similar to
our result.
Motsopoulos et al.\cite{Matsopoulos} for their vessel detection applied both morphological
processing and match filter. Their technique is disturbed by ONH edge and in their results there are some false vessels as well as some vessels are
missed.
Kochner et al \cite{Kochner} identified between 70-90\% of retinal vessel by applying tracking
and steerable filters. But their technique has some spurious vessels.
The methods based on morphological processing are for Zana and Klein \cite{Klein1, Klein2} that
had very good results in blood vessel segmentation in fluorescein angigrams images.
The negative point of their technique again is the detection of edge of ONH as a vessel
incorrectly.

Estabridis and De figueiredo reported a parallel algorithm to detect retinal blood
vessels, the algorithm was tested with 20 images, 10 normal and 10 abnormal and the
results demonstrate the robustness of the algorithm in the presence of noise. An
average true positive rate of 86.3\% with a false positive rate of 3.9\% is accomplished
with this algorithm when tested against hand-labeled data \cite{Estabridis}
Some of the automated techniques utilized fluorescein angiography and relied on
image-processing procedures to segment retinal blood vessel from the background
fundus image. Tavakoli et al. \cite{tavakoli-radon} worked on Fluorescein Angiography (FA) retinal image
by using Radon transform algorithm for detecting retinal vasculature. Fortunately their results
was good but because of invasive method of FA this technique can't use for
screening system. In other words, the drawbacks related to fluorescein angiography as
an intervention prohibit its use in large scale screening analysis.
However, their results similar this study was good enough and can use their method for
automated diagnostic system.
On other hand, Martinez-Perez et al. \cite{Martinez-Perez} presented an automated technique for segmenting
retinal blood vessels in FA retinal images and red-free based on multiscale feature
extraction. This technique overcomes the problem of inherent contrast variations in
FA images by applying the first and second spatial derivatives that gives information
about vessel topology. This approach also detects the blood vessels with different
widths, lengths and orientations. Comparison of their algorithm with first public
database yields values of 75.05\% true positive rate (TPR) and 4.38\% false positive
rate (FPR). Second public database values are of 72.46\% TPR and 3.45\% FPR. But their
results are not sensitive enough so as to evaluate the performance of vessel geometry
identification.
But the important problem of some studies \cite{LKalviainen, Tamura} were the need for interaction of
user as well as, huge measure of preprocessing computations.
As well as, another of the advantage of our algorithm to identify the blood vessel is
any requiring to location and identification of ONH  \cite{GoldbaumGlucom, Septiarini}.
Small vessels can appear as a succession of small isolated patterns, which could be
detected as false positive in detection of MA and/or HE. Vascular detection is the first
step of main processing, for DR detection.

One of the most important parts of in this study is to use multi overlapping window to
prevent error caused by global techniques. Applying this technique help use to find
small vessels and prevent from that problem said above as well as, this technique
caused incredible increasing in the efficiency of our algorithm.
One advantage of the automatic screening system is that it is deterministic, in other
words, it always classifies funduses in the similar way. There will always exist
differences in the backgrounds and education of human screeners, which causes
dispersion in their diagnose making. Also a single human expert may make different
diagnoses in different screening times due to human factors, such as tiredness or
sickness. A computer-based screening system does not have to be perfect to be used in
screening. It is better to use a computer-based screening system for classifying only
clearly normal funduses as normal, whereas abnormal and obscure funduses are
delivered to a human expert for further classification. However, the computer-based
screening system reduces the workload of the human expert, since in the screening
most of the funduses are normal, and only a few funduses have retinopathy. \\
Diabetes caused pathological alterations in the retinal blood vessel network that need
to detection and treatment as soon as possible in order to prevent DR.
For improving the evaluation of the retina condition image processing techniques are
required to investigate appropriate data about changes in retinal vasculature.
We divided our images into a training set with 60 images and test set with 220 images.
The training set was used for developing the algorithms and the test set only for
testing the algorithms.
Our results for vessel detection are incredible. The results proved that it is possible to
use algorithms for assisting an ophthalmologist to segment fundus images into normal
parts and lesions, and thus support the ophthalmologist in his or her decision making.
The algorithms detect regions where the image quality is inadequate, and thus it is
possible to show to the ophthalmologist what regions are left unprocessed.
In the screening it would be important that the computer-based system has very high
sensitivity.\\
The quality of our segmentation
depends on some parameters such as the length of our window (n), measure of step,
line validation thresholding, etc. determining of these parameter appropriately has
some advantages in our processing likes:\\
Accurate detection of retinal vessels location, Determination of some parameters like width and length of vessels, and Even determining of location of vessel bifurcation which can be assist to
clinician for analyzing image in later by registration scheme.\\
Our algorithm has some important characteristics in detection of vascular structure in
retinal images that include:\\
1. This algorithm is able to determine location, width and angle of vessels.\\
2. The algorithm is robust to noise.\\
3. Because of combination of two methods vessel segmentaion methods with the multi-overlapping window the performance of algorithm in detection of thick and
even thin vessels is acceptable.\\
Most of studies in this field are vessel detection based on morphological processing
and vessel tracking that in most of them ONH are key points of the algorithms.


%

\appendices


\ifCLASSOPTIONcaptionsoff
  \newpage
\fi


\begin{thebibliography}{1}


\bibitem{Khansari-TMI}
Khansari MM, Wanek J, Felder AE, Camardo N, Shahidi M. Automated assessment of hemodynamics in the conjunctival microvasculature network. IEEE Transactions on Medical Imaging. 2016 Feb;35(2):605-11.

\bibitem{Tavakoli-SPECT}
Tavakoli M, Najib M, Abdollahic A, Kalantarid F. Attenuation Correction in SPECT Images Using Attenuation Map Estimation with Its Emission Data. InSPIE Medical Imaging 2017 Mar 9 (pp. 101324Z-101324Z). International Society for Optics and Photonics.

\bibitem{TavakoliFA}
Tavakoli M, Mehdizadeh A, Pourreza R, Banaee T, Bahreyni Toossi MH, Pourreza HR. Early Detection of Diabetic Retinopathy in Fluorescent Angiography Retinal Images Using Image Processing Methods. Iranian Journal of Medical Physics. 2010;7(4):7-14.

\bibitem{Niemeijer1}
Niemeijer M, Abràmoff MD, Van Ginneken B. Fast detection of the optic disc and fovea in color fundus photographs. Medical image analysis. 2009 Dec 31;13(6):859-70.

\bibitem{sinthanayothin}
Sinthanayothin C, Boyce JF, Cook HL, Williamson TH. Automated localisation of the optic disc, fovea, and retinal blood vessels from digital colour fundus images. British Journal of Ophthalmology. 1999 Aug 1;83(8):902-10.

\bibitem{Mehdizadeh-color}
Pourreza HR, Bahreyni Toossi MH, Mehdizadeh A, Pourreza R, Tavakoli M. Automatic Detection of Microaneurysms in Color Fundus Images using a Local Radon Transform Method. Iranian Journal of Medical Physics. 2009 Mar 1;6(1):13-20.


\bibitem{Sherwani SM}
Sherwani SM, Tiwana MI, Iqbal J, Lovell NH. Automated Segmentation of Optic Disc Boundary and Diameter Calculation Using Fundus Imagery. InProceedings of the 2015 Seventh International Conference on Computational Intelligence, Modelling and Simulation 2015 Jul 27 (pp. 92-96). IEEE Computer Society.

\bibitem{Constante P}
Constante P, Gordón A, Chang O, Pruna E, Escobar I. Neural networks for optic nerve detection in digital optic fundus images. InAutomatica (ICA-ACCA), IEEE International Conference on 2016 Oct 19 (pp. 1-5). IEEE.

\bibitem{Khansari-DR2}
Khansari MM, O’Neill W, Penn R, Blair NP, Chau F, Shahidi M. An automated image processing method for classification of diabetic retinopathy stages from conjunctival microvasculature images. InSPIE Medical Imaging 2017 Mar 13 (pp. 101372C-101372C). International Society for Optics and Photonics.


\bibitem{Gagnon L}
Gagnon L, Lalonde M, Beaulieu M, Boucher MC. Procedure to detect anatomical structures in optical fundus images. InProc. SPIE 2001 Feb 19 (Vol. 4322, pp. 1218-1225).


\bibitem{Heneghan}
Heneghan C, Flynn J, O’Keefe M, Cahill M. Characterization of changes in blood vessel width and tortuosity in retinopathy of prematurity using image analysis. Medical image analysis. 2002 Dec 31;6(4):407-29.

\bibitem{Khansari-optics}
Khansari MM, O’Neill W, Penn R, Chau F, Blair NP, Shahidi M. Automated fine structure image analysis method for discrimination of diabetic retinopathy stage using conjunctival microvasculature images. Biomedical optics express. 2016 Jul 1;7(7):2597-606.

\bibitem{Fraz1}
Fraz MM, Remagnino P, Hoppe A, Uyyanonvara B, Rudnicka AR, Owen CG, Barman SA. Blood vessel segmentation methodologies in retinal images–a survey. Computer methods and programs in biomedicine. 2012 Oct 31;108(1):407-33.

\bibitem{Tavakoli-ONH1}
Pourreza-Shahri R, Tavakoli M, Kehtarnavaz N. Computationally efficient optic nerve head detection in retinal fundus images. Biomedical Signal Processing and Control. 2014 May 31;11:63-73.

\bibitem{tavakoli-twopreprocessingsteps}
Tavakoli M, Nazar M, Mehdizadeh A. Effect of Two Different Preprocessing Steps in Detection of Optic Nerve Head in Fundus Images. In SPIE Medical Imaging 2017 Mar 3 (pp. 101343A-101343A). International Society for Optics and Photonics.

\bibitem{Morales S}
Morales S, Naranjo V, Angulo J, Alcañiz M. Automatic detection of optic disc based on PCA and mathematical morphology. IEEE transactions on medical imaging. 2013 Apr;32(4):786-96.

\bibitem{Muangnak N}
Muangnak N, Aimmanee P, Makhanov S, Uyyanonvara B. Vessel transform for automatic optic disk detection in retinal images. IET Image Processing. 2015 Sep 1;9(9):743-50.


\bibitem{Abdullah M}
Abdullah M, Fraz MM, Barman SA. Localization and segmentation of optic disc in retinal images using circular Hough transform and grow-cut algorithm. PeerJ. 2016 May 10;4:e2003.



\bibitem{Sinthanayothin2}
Sinthanayothin C, Boyce JF, Williamson TH, Cook HL, Mensah E, Lal S, Usher D. Automated detection of diabetic retinopathy on digital fundus images. Diabetic medicine. 2002 Feb 1;19(2):105-12.
 
\bibitem{Hoover1}
Hoover AD, Kouznetsova V, Goldbaum M. Locating blood vessels in retinal images by piecewise threshold probing of a matched filter response. IEEE Transactions on Medical imaging. 2000 Mar;19(3):203-10.

\bibitem{Hoover2}
Hoover A, Goldbaum M. Locating the optic nerve in a retinal image using the fuzzy convergence of the blood vessels. IEEE transactions on medical imaging. 2003 Aug;22(8):951-8.


\bibitem{Staal}
Staal J, Abràmoff MD, Niemeijer M, Viergever MA, Van Ginneken B. Ridge-based vessel segmentation in color images of the retina. IEEE transactions on medical imaging. 2004 Apr;23(4):501-9.

\bibitem{Soares}
Soares JV, Leandro JJ, Cesar RM, Jelinek HF, Cree MJ. Retinal vessel segmentation using the 2-D Gabor wavelet and supervised classification. IEEE Transactions on medical Imaging. 2006 Sep;25(9):1214-22.


\bibitem{Jiang X}
Jiang X, Mojon D. Adaptive local thresholding by verification-based multithreshold probing with application to vessel detection in retinal images. IEEE Transactions on Pattern Analysis and Machine Intelligence. 2003 Jan;25(1):131-7.


\bibitem{Wu D}
Wu D, Zhang M, Liu JC, Bauman W. On the adaptive detection of blood vessels in retinal images. IEEE Transactions on Biomedical Engineering. 2006 Feb;53(2):341-3.

\bibitem{Tolias}
Tolias YA, Panas SM. A fuzzy vessel tracking algorithm for retinal images based on fuzzy clustering. IEEE Transactions on Medical Imaging. 1998 Apr;17(2):263-73.

\bibitem{Martinez-Perez2}
Martínez-Pérez ME, Hughes AD, Stanton AV, Thom SA, Bharath AA, Parker KH. Retinal blood vessel segmentation by means of scale-space analysis and region growing. InInternational Conference on Medical Image Computing and Computer-Assisted Intervention 1999 Sep 19 (pp. 90-97). Springer, Berlin, Heidelberg.

\bibitem{Zana}
Zana F, Klein JC. Segmentation of vessel-like patterns using mathematical morphology and curvature evaluation. IEEE transactions on image processing. 2001 Jul;10(7):1010-9.


\bibitem{Gardner}
Gardner GG, Keating D, Williamson TH, Elliott AT. Automatic detection of diabetic retinopathy using an artificial neural network: a screening tool. British journal of Ophthalmology. 1996 Nov 1;80(11):940-4.

\bibitem{Walter}
Walter T, Klein JC. Segmentation of color fundus images of the human retina: Detection of the optic disc and the vascular tree using morphological techniques. Medical data analysis. 2001:282-7.

\bibitem{Tavakoli-ONH2}
Tavakoli M, Toosi MB, Pourreza R, Banaee T, Pourreza HR. Automated optic nerve head detection in fluorescein angiography fundus images. InNuclear Science Symposium and Medical Imaging Conference (NSS/MIC), 2011 IEEE 2011 Oct 23 (pp. 3057-3060). IEEE.

\bibitem{Can}
Can A, Shen H, Turner JN, Tanenbaum HL, Roysam B. Rapid automated tracing and feature extraction from retinal fundus images using direct exploratory algorithms. IEEE Transactions on information Technology in Biomedicine. 1999 Jun;3(2):125-38.

\bibitem{Vlachos}
Vlachos M, Dermatas E. Multi-scale retinal vessel segmentation using line tracking. Computerized Medical Imaging and Graphics. 2010 Apr 30;34(3):213-27.

\bibitem{Goldbaum}
Goldbaum MH, Katz NP, Chaudhuri S, Nelson M. Image understanding for automated retinal diagnosis. InProceedings of the annual symposium on computer application in medical care 1989 Nov 8 (p. 756). American Medical Informatics Association.


\bibitem{tavakoli-fluoresceinangiography}
Tavakoli M, Shahri RP, Pourreza H, Mehdizadeh A, Banaee T, Toosi MH. A complementary method for automated detection of microaneurysms in fluorescein angiography fundus images to assess diabetic retinopathy. Pattern Recognition. 2013 Oct 31;46(10):2740-53.

\bibitem{tavakoli-radon}
Tavakoli M, Mehdizadeh AR, Pourreza R, Pourreza HR, Banaee T, Toosi MB. Radon transform technique for linear structures detection: application to vessel detection in fluorescein angiography fundus images. InNuclear Science Symposium and Medical Imaging Conference (NSS/MIC), 2011 IEEE 2011 Oct 23 (pp. 3051-3056). IEEE.

\bibitem{researchsection}
Research Section, Digital Retinal Image for Vessel Extraction (DRIVE) Database. Utrecht, The Netherlands, Univ. Med. Center Utrecht, Image Sci. Inst. [Online]. Available: http://www.isi.uu.nl/Re-search/Databases/DRIVE

\bibitem{MESSIDOR}
Decenciere E, Zhang X, Cazuguel G, Laÿ B, Cochener B, Trone C, Gain P, Ordonez R, Massin P, Erginay A, Charton B. Feedback on a publicly distributed image database: the Messidor database. Image Analysis and Stereology. 2014 Aug 26;33(3):231-4.

\bibitem{Lowell J}
Lowell J, Hunter A, Steel D, Basu A, Ryder R, Kennedy RL. Measurement of retinal vessel widths from fundus images based on 2-D modeling. IEEE transactions on medical imaging. 2004 Oct;23(10):1196-204.

\bibitem{Lalonde}
Lalonde M, Beaulieu M, Gagnon L. Fast and robust optic disc detection using pyramidal decomposition and Hausdorff-based template matching. IEEE transactions on medical imaging. 2001 Nov;20(11):1193-200.

\bibitem{Canny}
Canny J. A computational approach to edge detection. IEEE Transactions on pattern analysis and machine intelligence. 1986 Nov(6):679-98.

\bibitem{Zana}
Zana F, Klein JC. Segmentation of vessel-like patterns using mathematical morphology and curvature evaluation. IEEE transactions on image processing. 2001 Jul;10(7):1010-9.

\bibitem{Chaudhuri}
Chaudhuri S, Chatterjee S, Katz N, Nelson M, Goldbaum M. Detection of blood vessels in retinal images using two-dimensional matched filters. IEEE Transactions on medical imaging. 1989 Sep;8(3):263-9.



\bibitem{welikala}
Welikala RA, Dehmeshki J, Hoppe A, Tah V, Mann S, Williamson TH, Barman SA. Automated detection of proliferative diabetic retinopathy using a modified line operator and dual classification. Computer Methods and Programs in Biomedicine. 2014 May 31;114(3):247-61.

\bibitem{welikala2}
Welikala RA, Fraz MM, Dehmeshki J, Hoppe A, Tah V, Mann S, Williamson TH, Barman SA. Genetic algorithm based feature selection combined with dual classification for the automated detection of proliferative diabetic retinopathy. Computerized Medical Imaging and Graphics. 2015 Jul 31;43:64-77.


\bibitem{Khansari-DR}
Khansari MM, O’Neill W, Lim J, Shahidi M. Method for quantitative assessment of retinal vessel tortuosity in optical coherence tomography angiography applied to sickle cell retinopathy. Biomedical Optics Express. 2017 Aug 1;8(8):3796-806.

\bibitem{Kochner}
Kochner B, Schuhmann D, Michaelis M, Mann G, Englmeier KH. Course tracking and contour extraction of retinal vessels from color fundus photographs: Most efficient use of steerable filters for model-based image analysis. InProceedings of the SPIE Conference on Medical Imaging 1998 Feb (Vol. 3338, pp. 755-761).

\bibitem{Matsopoulos}
Matsopoulos GK, Mouravliansky NA, Delibasis KK, Nikita KS. Automatic retinal image registration scheme using global optimization techniques. IEEE Transactions on Information Technology in Biomedicine. 1999 Mar;3(1):47-60.

\bibitem{Martinez-Perez}
Martinez-Perez ME, Hughes AD, Thom SA, Bharath AA, Parker KH. Segmentation of blood vessels from red-free and fluorescein retinal images. Medical image analysis. 2007 Feb 28;11(1):47-61.

\bibitem{Klein1}
Zana F, Klein JC. Segmentation of vessel-like patterns using mathematical morphology and curvature evaluation. IEEE transactions on image processing. 2001 Jul;10(7):1010-9.

\bibitem{Klein2}
Zana F, Klein JC. A multimodal registration algorithm of eye fundus images using vessels detection and Hough transform. IEEE transactions on Medical Imaging. 1999 May;18(5):419-28.

\bibitem{Estabridis}
Estabridis K, Defigueiredo R. Blood vessel detection via a multi-window parameter transform. InComputer-Based Medical Systems, 2006. CBMS 2006. 19th IEEE International Symposium on 2006 Jun 22 (pp. 424-429). IEEE.


\bibitem{LKalviainen}
Kalviainen H, Hirvonen P, Xu L, Oja E. Probabilistic and non-probabilistic Hough transforms: overview and comparisons. Image and vision computing. 1995 May 1;13(4):239-52.

\bibitem{Tamura}
Tamura S, Okamoto Y, Yanashima K. Zero-crossing interval correction in tracing eye-fundus blood vessels. Pattern recognition. 1988 Jan 1;21(3):227-33.

\bibitem{GoldbaumGlucom}
Goldbaum MH, Sample PA, White H, Colt B, Raphaelian P, Fechtner RD, Weinreb RN. Interpretation of automated perimetry for glaucoma by neural network. Investigative ophthalmology and visual science. 1994 Aug 1;35(9):3362-73.

\bibitem{Septiarini}
Septiarini A, Harjoko A, Pulungan R, Ekantini R. Optic disc and cup segmentation by automatic thresholding with morphological operation for glaucoma evaluation. Signal, Image and Video Processing. 2017 Jul 1;11(5):945-52.







\end{thebibliography}
\end{document}